\documentclass{article}
\usepackage[utf8]{inputenc}
\usepackage[T1]{fontenc}
\usepackage{graphicx}

\usepackage[a4paper,total={7in, 10in}]{geometry}

\title{\textbf{Calculations of Particle Bombardment due to Dust and Charged Particles in the ISM on the Project Starshot Gram-Scale Interstellar Probe}}
\author{Kelvin F.Long\thanks{interstellarresearchcentre@gmail.com}
\\Interstellar Research Centre, Stellar Engines Ltd
\\100 Berkshire Place, Winnersh, Wokingham, 
\\ RG41 5RD, United Kingdom}

\begin{document}
\maketitle
\begin{abstract}

The Breakthrough Initiatives Project Starshot proposes to send a gram-scale laser driven spacecraft to the Alpha Centauri system in a 20 year mission travelling at $v \sim$ 0.2c. One of the challenges of this mission as the spacecraft moves through the interstellar medium is the presence of dust and gas (mostly hydrogen). The dust has a typical matter-density of $2.57 \times 10^{-27} gcm^{-3}$ with typical particle mass being $3 \times 10^{-13}$ g although some of the largest particles may be $5 \times10^{-9} g$ in mass. These dust particle will deposit $ \sim 10^{12}-10^{16}$ MeV onto the spacecraft with an energy flux of order $\sim 0.3 Js^{-1}m^{-2}$. We consider the erosion of the spacecraft frontal area due to dust and also heating effects. We attempt to characterise the likely environment for the starshot mission and estimate the particle bombardment shielding requirements in terms of mass and thickness of material. Current analysis estimates that the likely erosion rates are of order $ \sim 10^{-11}-10^{-8} gs^{-1}$ and that the frontal area temperature for the models examined in this paper will be $ \sim 135.2 K$ depending on the ratio of frontal area to radiating area. For an assumed shielding material with atomic number range $3-13$ (Lithium to Aluminium), and for spacecraft geometries with radii $\sim$1 mm and cylindrical length $\sim$5 mm, over a 21.5 year mission duration, this would suggest a shielding thickness of $ \sim$1.4 - 3 mm. This would also suggest a shielding mass in the range $\sim$0.01 - 0.05 g; depending on the material choice, spacecraft size and chosen geometry. This would represent between $\sim1 - 5\%$ of the total mass, assuming a spacecraft mass of 1g (driven by a $\sim10^2$ GW laser power). We also examine the additional effects of charged particles and estimate the stopping power and penetration range for different materials. Finally, we briefly examine the potential to use the incoming energy flux as a power source for the transmission of an optical laser deep space communication system. The work presented highlights the close coupling in the Project Starshot spacecraft design between the vehicle geometry and the particle bombardment requirements.

\end{abstract}


\section[Introduction]{Introduction}

The Breakthrough Initiatives Project Starshot \cite{breakthrough} is a private initiative to send a gram-scale interstellar spacecraft towards an exoplanet of a nearby star system within decades \cite{lubin}. It is also based on the US government funded Project Starlight under the NASA Innovative Advanced Concepts (NIAC) studies \cite{lubin1, lubin2}. It is proposed to send a spacecraft out through the solar heliosphere and out to the pristine interstellar medium (ISM) and beyond. 

The two Voyager spacecraft launched in 1977 have both now left the Solar System, but the interstellar probe proposes to go $\sim 10^3$ times this distance. The mission could be launched by the middle of this century and would involve using a ground based 100 GW power laser beam transmitted onto a $\sim$3 m sail craft in Earth orbit which contains the miniaturised payload. It would be pushed to a velocity of 60,000 km/s or 0.2$c$ (where $c$=299,792,458 m/s is the vacuum speed of light) in a matter of minutes, accelerating at over 10,000 g's. The goal would be to reach the mission target within around twenty years from launch, most likely headed to the Proxima Centauri star at 4.3 LY distance or around $\sim$260,000 AU. To put this into perspective, this would be equivalent to a cruise velocity of 12,650 AU/year, compared to the Voyager probes $\sim$3.5 AU/year. This is no small challenge, and the overarching research program of Project Starshot seeks to address all of the physics and engineering challenges that are on the critical pathway to success. In this paper, we focus on just one of these issues, which is the impact of interstellar dust particles on the spacecraft throughout the flight. 

As the spacecraft travels through the Solar System and eventually out to interstellar space, it is expected that it will encounter dust particles. In particular, it is known that interstellar dust particles exist, such as from the Stardust mission \cite{brownlee} and it is necessary to characterise the likely physics environment for the spacecraft. This is especially a concern since some of these particles may be ng in mass and as large as 0.1 $\mu$m in size, and so have the potential to deposit significant energy onto any impacting surface. To illustrate, if we assume a 1 $ng$ static dust particle impacting a surface which has a velocity of 100 km/s, it will have a kinetic energy on impact of 0.005 J or $\sim 3.1\times 10^{10}$ MeV, where 1eV is defined as the kinetic energy gained by one electron in falling through a potential difference of 1 volt.

In this paper, we describe the expected dust properties and calculate the likely effects on the interstellar probe mission so as to assess whether any shielding would be required at the front end. We also consider the physics penetration effects due to high energy electrons, protons, ions and cosmic rays. These need to be considered for the mission since around $\sim 99\%$ of the interstellar medium is gas and only $\sim 1\%$ is dust. Any spacecraft needs to mitigate against all of these effects if it is to survive the journey to a nearby star system. 

For all calculations (dust and charged particles) we assume that all impacts are incident normal to the absorber surface so that the path length of the incident particle can be assumed to be equal to the implant depth or range. This is also the path of maximum energy transfer when the collision is head-on. No account was made for the possibility of back-sputtering of incident particles due to recoil, the degree of which would increase with increasing angle of incidence. No account was made for ion channelling, whereby ions may penetrate greater distances along low index directions which may increase the range in these regions. 

Finally, we appreciate that the particle population density and distribution will be different in the interstellar medium from the interplanetary medium, and indeed also from the exoplanetary medium of the target star system. However, since we are only interested in a first assessment of the likely shielding requirements throughout the cruise phase of the flight, in this work we only consider the properties of the interstellar medium of interest. This is justified on the basis of two reasons (1) Given the high acceleration of the probe during the laser beaming phase, it will quickly get to the edge of the known Solar System and in any case characterising the interplanetary medium dust and particle properties should be much easier than for further distances (2) If calculations show that getting through the interstellar medium will prove difficult, then the properties of the exoplanetary medium become purely academic.

\hspace*{5in}

\section[Model Calculations \& Analysis]{Model Calculations \& Analysis}

\subsection[Impact Events]{Impact Events}

We start by making some assumptions about the mission. The mission is to be completed in around 20 years. The spacecraft would be exposed to the interstellar dust particles at its peak velocity for less time than this. However, for the purposes of being conservative and allowing for margins and uncertainties we assume exposure to dust throughout the linear flight time. For the model that was run in this work, the actual mission time was 21.5 years.

We next make some assumptions about the properties of the interstellar medium environment that we can expect to fly through. Evidence for gas and dust grain properties comes from the space missions. This includes HEOS-2 (1972-1974), Pioneer 10 (1972-2003), Pioneer 11 (1973-1995), Voyager 1 and 2 (1977-present), Galileo (1989-2003), Ulysses (1990-2009), Cassini (1997-2007), Helios (1974-1985), Giotto (1985-1992), Proba 1 and 2 (2001-2016), Rosetta (2004-2016), New Horizons (2006-present). For example, the Galileo mission measured Dust mass range $10^{-6}-10^{-7}$ g, speed 1-70 km/s, and a mean particle mass of $2 \times 10^{-12}$ g. The Ulysses mission measured a mean particle mass of $1 \times 10^{-12}$ g. For the Voyager 1 and 2 missions the dust particle size did not exceed 1 $\mu$m at 50 AU \cite{landis}. From what we know today dust particles are typically around 1-50 $\mu$m at 50 AU. The latest estimates \cite{draine, draine1} suggest that the matter density is believed to be around $\rho = 2.57 \times 10^{-27}$ gcm\textsuperscript{-3} and this is similar to modern estimates reported by other authors \cite{lubin, landis, crawford, early, hoang}.

We model the impact as a cylindrical spacecraft incident upon a stationary dust particle of mass $m_d$. Such a dust particle would then impact somewhere on the spacecraft front face and trace a column of depth $t$ along the length of the spacecraft, with the column mass given by:

\begin{equation}
m_s=\rho\pi r^2 t
\end{equation}

Where $\rho$ is the shield material density and $r$ is the impact column radius. This model is valid provided that the mass of the dust particle exceeds the column mass of the spacecraft $m_d>m_s$. We conduct calculations for shield materials constructed of Lithium ($\rho=530$ g/cm\textsuperscript{3}), Beryllium ($\rho=1850$ g/cm\textsuperscript{3}), Boron ($\rho=2500$ g/cm\textsuperscript{3}), Graphite ($\rho=2300$ g/cm\textsuperscript{3}) and Aluminium ($\rho=2700$ g/cm\textsuperscript{3}). Tables 1-2 show an estimate for the column thickness as a function of material to illustrate how the estimated column mass changes with assumed column depth. Next, assuming an impacting column with radius $r=$ 400 nm and a column depth $t=$100 nm, with the material constructed of Aluminium, we calculate that the mass of the impacting dust column will be $1.35\times 10^{-13}$ g. 

\begin{table}[ht!]
\begin{center}
     \begin{tabular}{| p{3cm} | p{2.2cm} | p{2.2cm}| p{2.2cm}| p{2.2cm}|} \hline
    \textbf{Material} & \textbf{$t$ = 1 nm}  & \textbf{$t$ = 10 nm} & \textbf{$t$ = 100 nm} & \textbf{$t$ = 1,000 nm}  \\ \hline
    Lithium &  $2.7\times10^{-16}$  & $2.7\times10^{-15}$ & $2.7\times10^{-14}$  & $2.7\times10^{-13}$   \\ \hline
    Beryllium  & $9.3\times10^{-16}$ & $9.3\times10^{-15}$ & $9.3\times10^{-14}$ & $9.3\times10^{-13}$   \\ \hline
Boron  & $1.3\times10^{-15}$ & $1.3\times10^{-14}$ & $1.3\times10^{-13}$ & $1.3\times10^{-12}$  \\ \hline
Graphite  & $1.2\times10^{-15}$ & $1.2\times10^{-14}$ & $1.2\times10^{-13}$ & $1.2\times10^{-12}$   \\ \hline
Aluminium  & $1.4\times10^{-15}$ & $1.4\times10^{-14}$ & $1.4\times10^{-13}$ & $1.4\times10^{-12}$  \\ \hline
            \end{tabular}
     \caption{\textbf{Estimates of column mass in grams for different spacecraft materials and for various column depths, assuming a 400 nm diameter impacting dust particle.}}
\end{center}
 \end{table}

\begin{table}[ht!]
\begin{center}
     \begin{tabular}{| p{3cm} | p{2.5cm} | p{2.5cm}| p{2.5cm}| } \hline
    \textbf{Material} & \textbf{Typical Mass} ($3\times10^{-13}$ g)  & \textbf{Mean Mass} ($2\times10^{-12}$ g) & \textbf{Large Mass} ($1\times10^{-9}$ g)   \\ \hline
    Lithium &  1,126  & 7,507 &  $3.75\times10^{6}$   \\ \hline
    Beryllium  & 322  & 2,151 & $1.07\times10^{6}$    \\ \hline
Boron  & 239 & 1,591 & $7.96\times10^{5}$   \\ \hline
Graphite  & 259  & 1,730 & $8.65\times10^{5}$    \\ \hline
Aluminium  & 221 & 1,474  & $7.36\times10^{5}$     \\ \hline
            \end{tabular}
     \caption{\textbf{Estimates of column thickness in nanometers for a $400 nm$ sized impacting dust particle.}}
\end{center}
 \end{table}

We can then calculate the kinetic energy $E$ of the impacting column mass, which we do for the different material densities and column thickness but assuming the Project Starshot cruise speed $v$ of $0.2c$ or 60,000 km/s. The kinetic energy is given by:

\begin{equation}
E=\frac{1}{2}\rho \pi r^2 t v^2
\end{equation}

The results of these calculations are shown in Table 3 and they range from $\sim 5\times10^{-4} - 2.4$ J. For interest, we can also then convert this to useful units of energy density as an indication of the potential energetic damage of these impacts, by dividing this energy by the mass of the column. As shown in Tables 4 - 6, we find this varies from the low end at 500 MJ/kg (lithium target 1nm column thickness, largest dust grain $1\times10^{-9}$g), to $8.143\times10^{9}$ MJ/kg (Aluminium target 1,000 nm column thickness, typical dust grain $3\times10^{-13}$g). 

\begin{table}[hp]
\begin{center}
      \begin{tabular}{| p{3cm} | p{2cm} | p{2cm}| p{2cm}| p{2cm}|} \hline
    \textbf{Material} & \textbf{1 nm}  & \textbf{10 nm} & \textbf{100 nm} & \textbf{1000 nm}  \\ \hline
    Lithium & 0.0005   & 0.0048  & 0.0479  &  0.4795   \\ \hline
    Beryllium  & 0.0017  & 0.0167  & 0.1674  & 1.6738    \\ \hline
Boron  & 0.0023 & 0.0226  & 0.2262  & 2.2619   \\ \hline
Graphite  & 0.0021 & 0.0208  & 0.2081  & 2.0810    \\ \hline
Aluminium  & 0.0024 & 0.0244  & 0.2443  & 2.4429   \\ \hline
            \end{tabular}
          \caption{\textbf{Estimates of kinetic energy (J) for different column mass assuming a 400 nm sized impacting dust particle.}}
\end{center}
 \end{table}

\begin{table}[ht!]
\begin{center}
     \begin{tabular}{| p{3cm} | p{2cm} | p{2cm}| p{2cm}| p{2cm}|} \hline
    \textbf{Material} & \textbf{1 nm}  & \textbf{10 nm} & \textbf{100 nm} & \textbf{1000 nm}  \\ \hline
    Lithium &  1.59 & 15.98  & 159.84  &  1,598.4  \\ \hline
    Beryllium  & 5.58 & 55.79  & 557.95 &  5,579.5  \\ \hline
Boron  & 7.54 & 75.40  & 753.98  &  7,539.8 \\ \hline
Graphite  & 6.94& 69.37 & 693.7 & 6,936.6   \\ \hline
Aluminium  & 8.14 & 81.43 & 814.3 & 8,143  \\ \hline
            \end{tabular}
     \caption{\textbf{Estimates of energy density (million MJ/kg) for different spacecraft materials assuming a typical dust particle mass of $3\times10^{-13}$ g.}}
\end{center}
 \end{table}

\begin{table}[ht!]
\begin{center}
     \begin{tabular}{| p{3cm} | p{2cm} | p{2cm}| p{2cm}| p{2cm}|} \hline
    \textbf{Material} & \textbf{1 nm}  & \textbf{10 nm} & \textbf{100 nm} & \textbf{1000 nm}  \\ \hline
    Lithium &  0.2398  & 2.397 & 23.98  & 239.8    \\ \hline
    Beryllium  & 0.8369 & 8.3692  & 83.69 & 836.9   \\ \hline
Boron  & 1.1309  & 11.3097  & 113.1  & 1,130.9  \\ \hline
Graphite  & 1.0405 & 10.4049  & 104.0  &  1,040.5  \\ \hline
Aluminium  & 1.2215 & 12.2145 & 122.1  &  1,221.5 \\ \hline
            \end{tabular}
     \caption{\textbf{Estimates of energy density (million MJ/kg) for different spacecraft materials assuming a mean dust particle mass of $2\times10^{-12}$ g.}}
\end{center}
 \end{table}

\begin{table}[ht!]
\begin{center}
     \begin{tabular}{| p{3cm} | p{2cm} | p{2cm}| p{2cm}| p{2cm}|} \hline
    \textbf{Material} & \textbf{1 nm}  & \textbf{10 nm} & \textbf{100 nm} & \textbf{1000 nm}  \\ \hline
    Lithium &  0.0005  & 0.0048 &  0.0479 & 0.4795   \\ \hline
    Beryllium  & 0.0017  & 0.0167 & 0.1674  &  1.6738  \\ \hline
Boron  & 0.0023 & 0.0226 & 0.2262 & 2.2619  \\ \hline
Graphite  & 0.0021 & 0.0208  & 0.2081  &  2.0810  \\ \hline
Aluminium  & 0.0024 & 0.0244 & 0.2443  & 2.4429  \\ \hline
            \end{tabular}
     \caption{\textbf{Estimates of energy density (million MJ/kg) for different spacecraft materials assuming a large dust particle mass of $1\times10^{-9}$ g.}}
\end{center}
 \end{table}

The dust in the interstellar medium is expected to have a typical solid density of around 3 g/cm\textsuperscript{3} and is mostly silicate with large fractions of carbonaceous material. Some of the largest dust grains may be as large as a $\sim$few ng \cite{draine}. In general, we can split the interstellar dust grains down into three distributions of particles masses. These are given below, along with the estimated number densities ($n= \rho/m$) from the matter density quoted above.

\begin{itemize}
\item Typical dust particles: 	$m = 3 \times 10^{-13}$ g, $n = 8.56 \times 10^{-15}$ cm\textsuperscript{-3}
\item Mean dust particles: 	          $m = 2 \times 10^{-12}$ g, $n = 1.28 \times 10^{-15}$ cm\textsuperscript{-3}
\item Largest dust particles: 	$m = 1 \times 10^{-9}$ g, $n = 2.57 \times 10^{-18}$ cm\textsuperscript{-3}
\end{itemize}

This means that for a 4.3 $ly$ mission target, we can multiply the number density $n$ by the mission distance in light years to get an estimate for the number of impact events per area per light year $N_{ly}$, or the number of impact events per area throughout the entire mission $N_{tot}$. This leads to the following estimates:

\begin{itemize}
\item Typical dust particles:	$N_{ly} \sim 8,100$ impacts/cm\textsuperscript{2}/LY, $N_{tot} \sim 34,800$ impacts/cm\textsuperscript{2} $\equiv$ 34 /mm\textsuperscript{2}
\item Mean dust particles:	           $N_{ly} \sim 1,200$ impacts/cm\textsuperscript{2}/LY, $N_{tot} \sim 5,200$ impacts/cm\textsuperscript{2} $\equiv$ 52 /mm\textsuperscript{2}
\item Largest dust particles:	$N_{ly} \sim 2$ impacts/cm\textsuperscript{2}/LY, $N_{tot} \sim 10$ impacts/cm\textsuperscript{2} $\equiv$ 0.1 /mm\textsuperscript{2}
\end{itemize}

We can also estimate the total number of impacts $N$, by multipying by the cross section area (we assume $r=1mm$ and $A_o$=3.14mm\textsuperscript{2} as described in section 3), and then estimate the impact frequency $t_{i}$ by diving by the total mission time which is assumed to be 20 years as follows:

\begin{itemize}
\item Typical dust particles: $N \sim107$ impacts, $t_{i} \sim1.69 \times10^{-7}$ impacts/s $\equiv 0.17 ~\mu$Hz
\item Mean dust particles:	  $N \sim164$ impacts, $t_{i} \sim2.59 \times10^{-7}$ impacts/s $\equiv 0.26 ~\mu$Hz
\item Largest dust particles: $N \sim0.31$ impacts, $t_{i} \sim4.98 \times10^{-10}$ impacts/s $\equiv 0.0004 ~\mu$Hz
\end{itemize}

As can be seen, the numbers show that statistically the number of dust impacts may be manageable throughout the flight, although each one has the potential to result in a critical failure due to its large kinetic energy and is analogous to firing a rifle bullet at the spacecraft.

We can also calculate the energy of these dust particles and then by multiplying this by the total number of impacts we can find the total energy deposition expected throughout the mission as they impact the spacecraft. This is velocity dependent and for the interstellar probe we assume 60,000 km/s. The energy deposition for this particle distribution, assuming fractions of particle mass, are calculated to be:

\begin{itemize}
\item Typical dust particles:	$E_k$ = 0.5 J $\equiv 3.4 \times 10^{12}$ MeV, $E_{tot}$ = 17.4 kJ/cm\textsuperscript{2}
\item Mean dust particles:	          $E_k$ = 3.6 J $\equiv 2.2 \times 10^{13}$ MeV, $E_{tot}$ = 18.7 kJ/cm\textsuperscript{2}
\item Largest dust particles:	$E_k$ = 1,800 J $\equiv 1.1 \times 10^{16}$ MeV, $E_{tot}$ = 18 kJ/cm\textsuperscript{2}
\end{itemize}

\subsection[Surface Temperature]{Surface Temperature}

The temperature of the frontal material becomes an issue when it rises to the point of material vaporisation. It is therefore necessary to conduct some analysis of the expected temperatures on the front of the spacecraft. We assume a model where energy is impacting the frontal area of the spacecraft, due to electrons and protons, and some of this is then converted into heat. As described in \cite{martin} the temperature rise due to heat generation is given by:

\begin{equation}
T^4=\frac{\epsilon A_o \phi}{A \sigma E_m}
\end{equation}

Where $A_o$ is the  projected frontal cross-sectional areas of the vehicle, $A$ is an assumed uniformly radiating spacecraft surface area, $\phi$ is the flux of energy into the vehicle (also known as \textit{fluence}), $\sigma$ is the Stefan-Boltzmann constant, $5.67 \times 10^{-8}$ Wm\textsuperscript{-2}K\textsuperscript{-4}, and $E_m$ is the surface emissivity. 

It was also shown by \cite{martin} that the energy flux into the vehicle has a dependence on the fractional velocity relative to the speed of light $\beta=v/c$, and the matter density of the incoming particle stream, $\rho=nm$ as follows:

\begin{equation}
\phi=\frac{\rho \beta c^{3}}{(1-\beta^{2})^{1/2}} \times \Big[\frac{1}{(1-\beta^{2})^{1/2}}-1 \Big]
\end{equation}

For the starshot mission, the spacecraft is travelling at 60,000 km/s ($\sim$12,650 AU/Year) which relative to the speed of light, $\sim 3 \times10^8$ m/s, is $\beta= 0.2$. We do therefore expect some relativistic effects for such a high speed mission. The calculated value for the energy flux is $\sim0.29$ J/sm\textsuperscript{2}.

Table 7 shows the results of calculating the surface temperature for different spacecraft aspect ratios. It can be seen that the surface temperature rises moderately with aspect ratio. The aspect ratio effectively represents the ratio between the bombarded and radiative surface areas. For the case where $A_o/A <1$ this represents a fully radiating surface. For the case where $A_o/A=A$  all incoming energy is re-radiated. For the case of $A_o/A>1$ only a fractional area of the surface is radiating the incoming energy, and so that heat will be retained within the spacecraft.

In the calculations we have assumed that the emissivity $E_m=1.0$ but if it is less than unity this will result in a raising of the surface temperature. For example a value of $E_m=0.8$ would result in a rise in the surface temperature by around $\sim5\%$.

To ensure that the chosen shield material does not go into the vaporisation regime (discussed in the next section) it would be necessary to choose materials which have sublimation temperatures below those indicated in Table 7, depending on the area ratio and the cruise velocity of the flight. 

\begin{table}[ht!]
\begin{center}
     \begin{tabular}{| p{3cm} | p{2cm} | p{2cm}| p{2cm}| p{2cm}|} \hline
    \textbf{Aspect Ratio $A_o/A$} & \textbf{$T_s$ (K) (0.2c)}   \\ \hline
    0.1 & 135.2     \\ \hline
    1.0  & 240.4   \\ \hline
10.0  & 427.5  \\ \hline
100.0  & 760.2   \\ \hline
1000.0  & 1,351.9  \\ \hline
            \end{tabular}
     \caption{\textbf{Surface Temperature due to protons and electrons for the Breakthrough Starshot spacecraft based on the ratio of frontal surface area $A_o$ to radiating surface area $A$ with an energy flux of 0.29 J/sm\textsuperscript{2}.}}
\end{center}
 \end{table}

For the design geometries discussed in the next section these have an aspect ratio of $A_o/A=0.1$ and so have a frontal surface temperature of 135.2 K. This is not too dissimilar to the temperature estimate of $\sim$100 K provided by Lubin in earlier work \cite{lubin}, with an assumed 'wafer' frontal surface radii of around $\sim$80 mm and area $\sim$0.02 m$^2$. It is also interesting that this same author discusses the case of long cylindrical geometry with reduced effective cross section to minimize dust hits and decrease frontal temperatures.

\subsection[Erosion Rates]{Erosion Rates}

The ablation rate or mass loss per unit time is derived by Benedikt \cite{benedikt} and also used by \cite{martin}. It is given by:

\begin{equation}
\frac{dm}{dt}= \frac{\eta A_o}{H_s}  \frac{\rho \beta c^3}{(1-\beta^{2})^{1/2}}  \Big[{\frac{1}{(1-\beta^{1/2})^{1/2}} - 1}\Big]
\end{equation}

Where $H_s$ is the latent heat of sublimation, $\eta$ is the fraction of energy which is transferred from the medium and results in permanent material changes, $\beta$ is the relativistic velocity factor. For the use of other authors who may wish to repeat some of the analysis here, there are other equations that can be used to estimate the erosion rates, such as the equation derived by Langton \cite{langton}:

\begin{equation}
\frac{dm}{dt}=\frac{A_o}{(S \Delta T + L)} \frac{\rho \beta^3 c^3}{2(1-\beta^2)^{1/2}}
\end{equation}

Where $S$ is the specific heat of the material, $\Delta T$ is the temperature change required to vaporise the metal and $L$ is the latent heat, which is equal to the sum of the latent heats of fusion and evaporation. However, in this work we restrict our analysis to the equation derived by Benedikt on the basis that other authors \cite{martin} state that Langton leads to incorrect values of the energy flow and higher values of the mass loss per unit time.

For the model of the starshot probe we assume a spacecraft cruise velocity of $\beta=v/c=0.2$ travelling to a distance of 4.3 LY ($\sim$272,000 AU). We assume a circular frontal surface geometry with a cylindrical body. We consider two possible geometries for this which we term '\textit{Tom}' and '\textit{Jerry}' (after the cartoon characters) with defined radius ($R$) and length ($l$), which have an aspect ratio of $A_o/A=0.1$. The geometries are as follows:
\begin{itemize}
\item \textit{Jerry}: $R$ = 1 mm, $l$ = 5 mm, $A_o$= 3.14 mm\textsuperscript{2}, $A$= 31.41 mm\textsuperscript{2}
\item \textit{Tom}: $R$ = 10 mm, $l$ = 50 mm, $A_o$= 314.159 mm\textsuperscript{2}, $A$= 3,141.59 mm\textsuperscript{2}
\end{itemize}

These are assuming cylindrical geometries where the frontal surface area is $\pi R^2$, and the radiating surface area is $(2\pi R^2)+l(2 \pi R)$ but with the end surface areas subtracted. This way the radiating surface area is only assumed to be the cylindrical circumferential area. It is assumed in these configurations that the sail had been ejected after the acceleration phase of the mission hence in this analysis we have separated the payload and the sail. These geometries are chosen as being representative boundary conditions for what a Starshot probe size might be. Anything smaller than a 1 mm frontal radius is likely difficult to manufacture, and anything with a radius larger than 10 mm would also imply a significant mass increase of the probe beyond the Gram-scale limit requirement. The cylindrical geometry may also permit for the isotropic radiation of any heat from its surface. Other geometries are also possible (square, circular, sphere, torus) but these can be studied at a later date since we are only interested in estimates of the likely erosion regime in this preliminary work. The properties of the interstellar medium dust are as stated earlier in this paper.

\begin{table}[ht!]
\begin{center}
     \begin{tabular}{| p{3cm} | p{2cm} | p{2cm}| p{2cm}| p{2cm}|} \hline
    \textbf{Material} & {Mass Loss (g/s)}  & {Shield Mass (g)} & {Shield Thickness (mm)}\\ \hline
    Lithium $z=3$ & $3.56\times10^{-10}$ & 0.241 & 145.09     \\ \hline
    Beryllium $z=4$  & $2.58\times10^{-11}$ & 0.017 & 3.01    \\ \hline
Boron $z=5$  & $1.71\times10^{-11}$ & 0.011  & 1.47  \\ \hline
Graphite $z=6$  & $1.53\times10^{-11}$ & 0.010  & 1.43   \\ \hline
Aluminium $z=13$  & $7.57\times10^{-11}$ & 0.051 & 1.62   \\ \hline
            \end{tabular}
     \caption{\textbf{Calculations ('Jerry' $R$ = 1 mm model) of Minimum Shielding Requirements for the Starshot probe over a 21.5 year flight at 60,000 km/s.}}
\end{center}
 \end{table}

\begin{table}[ht!]
\begin{center}
     \begin{tabular}{| p{3cm} | p{2cm} | p{2cm}| p{2cm}| p{2cm}|} \hline
    \textbf{Material} & {Mass Loss (g/s)}  & {Shield Mass (g)} & {Shield Thickness (mm)}\\ \hline
    Lithium $z=3$ & $3.56\times10^{-8}$ & 24.157 & 145.09     \\ \hline
    Beryllium $z=4$  & $2.58\times10^{-9}$ & 1.747 & 3.01    \\ \hline
Boron $z=5$  & $1.71\times10^{-9}$ & 1.158  & 1.47  \\ \hline
Graphite $z=6$  & $1.53\times10^{-9}$ & 1.035  & 1.43   \\ \hline
Aluminium $z=13$  & $7.57\times10^{-9}$ & 5.131 & 1.62   \\ \hline
            \end{tabular}
     \caption{\textbf{Calculations ('Tom' $R$ = 10 mm model) of Minimum Shielding Requirements for the Starshot probe over a 21.5 year flight at 60,000 km/s.}}
\end{center}
 \end{table}

The results of these calculations are shown in Tables 8 and 9. It can be seen that for the range of materials modelled, the mass loss over the flight does not exceed $\sim 4\times10^{-10}$ g/s for 'Jerry' and $\sim 4\times10^{-8}$ g/s for 'Tom'. However it is also clear that Lithium (with its low density) leads to too high a shield thickness, exceeding $\sim$100 mm, and Aluminium (with its high density) leads to too high a shielding mass, exceeding $\sim$5 g, for spacecraft designs with a larger cross sectional area than shown in the model for 'Jerry'. Materials with atomic numbers inside of these two extremes would therefore seem to be suitable. That said, there are other materials that can also be considered such as with Z>13 (particularly due to their increased stopping power for incident charge particles). Such possibilities have been explored by others \cite{lubin3} but were not considered in this study when the original calculations were concluded.

Taking account of dust only, for the slim 'Jerry' model and assuming a total spacecraft mass of 1 g, this suggests any shielding mass requirement will be in the range $\sim$0.01 - 0.05 g and with a thickness in the range $\sim$1.4 - 3 mm. For a starshot probe with a total mass of 1 g, this would constitute approximately $\sim$1 - 5\% of the total vehicle mass, depending on the spacecraft size and geometry. 

Similarly, for the 'Tom' model and assuming a 10 gram total spacecraft mass, this suggests any shielding mass requirement will be in the range $\sim$1 - 5 g and with a thickness in the range $\sim$1.4 - 3 mm. This would constitute approximately $\sim10 - 50\%$ of the total vehicle mass, and is clearly not a recommended design reference range given the Gram-scale requirement of Project Starshot. This suggests that the spacecraft frontal cross sectional radius should be $\ll$10 mm for the minimisation of shielding mass.

It is noted that the derived spacecraft thicknesses of $\sim$1.4 - 3 mm are consistent with the estimates made by Hoang \cite{hoang} assuming beryllium and graphite based shielding materials who recommended a requirement for $\sim$1 - 3 mm for the Starshot probe. That two separate studies, using different methods of analysis, arrive at similar numbers, gives confidence that we are correctly estimating the likely particle bombardment effects on the design.

Figure 1 shows an illustration of minimum starshot bombardment shielding set-up might look like for an assumed cylindrical geometry of 1 mm radii. What has not been considered in this work and also not illustrated in the figure is the additional need for margins and uncertainties on any defined engineering requirements.

It is likely that a 10 gram spacecraft is the outer limit on the design mass since the power of the propulsion beamer is a linear function of the mass, $P=mca/2\mu$, where $m$ is the spacecraft mass, $c$ is the speed of light in a vacuum, $a$ is the spacecraft acceleration, $\mu$ is the sail loading \cite{longbook}. Therefore, assuming the same design parameters, a $\sim 10^2$ GW laser propulsion power for a 1 gram starshot probe would turn into a $\sim 10^3$ GW beamer for a 10 gram probe in terms of order of magnitude estimates. 

Although the calculations suggest that the erosion of the front of the spacecraft throughout its flight is likely to be minimal, one must consider the possibility that a large dust particle of order 5 ng did impact the spacecraft at 100 km/s. This would have a kinetic energy on impact of order 0.005 J or equivalent to $3.1\times10^{10}$ MeV. What physics effects can we expect from such an impact? The same particle impacting at 60,000 km/s would have an energy of 9,000 J or $5.6\times10^{16}$ MeV.

If the particle was moving slower than the sound speed of the material surface that it impacts, then the dust particle would deposit its kinetic energy into the material and merely cause some material damage. However, if the dust particle is moving much faster than the sound speed of the material surface that it impacts, then this will create a small volume in the surface bounded by a shock front as the dust particle slows down to subsonic speeds. It will deposit its kinetic energy into that volume producing an increase in energy density in that region. This may result in an increase in local surface temperature or in a worst case scenario some material melting and permanent deformation of the material or ablative mass loss.

\begin{figure}[htb]
\includegraphics[width=5in]{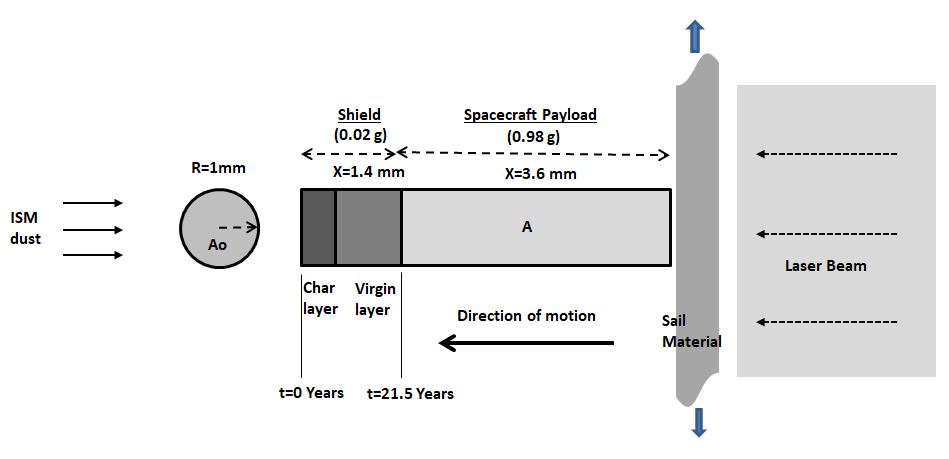}
\centering
\caption{\textbf{Illustration of minimum starshot particle bombardment shielding model for an assumed cylindrical geometry taking account of dust only.}}
\end{figure}

\subsection[Charged Particles]{Charged Particles}

Although we have calculated the erosion rate and therefore expected shield requirements due to dust impacts, another consideration is the penetration of energetic charged particles. When a charged particle penetrates matter, it will interact with the electrons in the absorbing target material via the Coulomb force, and so can cause the atoms in the target to undergo excitation or ionisation. As the particle enters the matter, momentum is transferred via Coulomb interactions and so the particle will slow down, but this interaction time is obviously a function of the particle speed. A faster particle will interact less with other particles that it passes. 

One of the sources of these energetic charged particles is Cosmic rays which originate outside of the Solar System, such as from supernova explosions and from distant galaxies such as Active Galactic Nuclei. The nature of these particles is mostly high energy protons and other atomic nuclei. Of those in deep space, approximately $\sim99\%$ are atoms that have had their electrons stripped off and around $\sim1\%$ are free electrons. Then from this approximately $\sim90\%$ are protons (hydrogen), $\sim9\%$ are \textsuperscript{4}He alpha particles, and $<1\%$ are other heavier nuclei such as Carbon, Oxygen, Magnesium, Silicon and Iron \cite{dembinski}. A very small amount of the cosmic rays may also be composed of antimatter particles, such as positrons (anti-electrons) and anti-protons). 

The interaction of Cosmic rays with a space probe can cause devastating effects, such as flip switching of logic gates in electronic components. It is likely that such an event may have disrupted the Voyager 2 spacecraft on the 22nd April 2010 when the spacecraft experienced scientific data format problems \cite{globus, antczak}. Crudely speaking, if after being in deep space a couple of decades, and approaching $\sim$100 AU the Voyager 2 spacecraft can experience 1 flip switching event from a Cosmic ray, then for a mission over a similar time span that goes $\sim10^3$ order of magnitude the distance (but much faster), it would not be unreasonable to expect of order $\sim10^3$ potential flip switching events awaiting a starshot probe as it heads towards its interstellar goal several light years away.

Considering protons (ionised hydrogen), if the speed of the impact is less than around $\sim$0.1c then the penetration effects can be treated as insignificant as an approximation. If however they exceed $\sim$0.1c then they will become significant \cite{martin}. If the speed of the impact exceeds greater than around $\sim$0.9c then the incoming protons will collide with the nuclei in the absorbing material causing atomic displacement, plastic deformation, nuclear disintegration and even local melting and permanent damage. If the impact speed is a lot less than this however (as it is for Starshot at $\sim$0.2c) then energy will be dissipated by interaction with the electrons in the absorber resulting in electron excitation and de-excitation as well as atomic ionisation and recombination. This will lead to radiation emission, absorption and heating of the material, and is the main reason for why the frontal area of the spacecraft gets hot.

In this work, we do not assess the effects of impacting electrons but it is worth mentioning the likely physics effects. The interaction of incoming electrons near any nuclei in the absorber will result in energy loss and radiation emission. If the impacting speed is approaching the speed of light, then this can result in pair formation and permanent damage to the material. At impacting speeds less than this but exceeding $\sim$0.8c one can expect Compton scattering; where energy is transferred from the electron to photons, resulting in the scattered photons having a higher frequency.  For impacting speeds less than $\sim$0.8c the absorbing material will undergo photoelectric effect; causing x-ray emission from the incoming electrons which causes heating of the material. In addition, there will be electron-electron interactions in addition to Bremsstrahlung radiation emission from the deceleration of the charged particle into the absorber material, which causes further absorption and heating. 

Another effect of charged particles for which the Starshot design team should be aware, but for which we do not calculate, is the effect of having a transitional layer in any shielding materials, or between the shielding layer of the spacecraft and the internal payload layer. When crossing boundaries of two different materials, both with different dielectrical constants, a charged particle will emit electromagnetic radiation, with an energy proportional to the Lorentz factor $\gamma$. This is a concern particularly for any charged particles that have a range that exceeds the depth of the shielding absorber, and may generate radiation internal to the spacecraft \cite{jackson}.

We do not intend here to complete a comprehensive analysis of the charged particle effects, but merely to get a feel for the additional requirements they place on the spacecraft design. Eventually, for the Project Starshot spacecraft design, it will be necessary to take account of all of these effects to properly characterise the likely physics environment and therefore the shielding requirements to protect the sensitive payload.

During the Project Daedalus particle bombardment study \cite{martin}, the authors looked briefly at the stopping of interstellar ions, but they only looked at 10 MeV protons, concluding that for Beryllium the penetration distance would be around 0.8 mm, suggesting that the estimated particle shield thickness of 9 mm would be approximately 10 times the proton range and so the transmission will be essentially non-existent. 

Table 10 shows the collision energy of different elements in the interstellar medium, which we assume is dominated by hydrogen and helium (making up $\sim$93\% and $\sim$7\% respectively), in addition to small traces of other elements like Carbon and Oxygen. We neglect relativistic effects for this illustration. As can be seen the collision energy increases with the element mass and also with the collision speed.

The collisional impact energy due to hydrogen and helium, are the two dominant constitutes of the interstellar medium, making up around ~93\% and ~7\% respectively. The table also shows the impact energy due to Oxygen and Carbon, representative of the ~0.1\% ‘metal’s found in the interstellar medium. For a spacecraft speed of $0.2c$ the expected impact energies for hydrogen and helium are $\sim$18.8 MeV and $\sim$74.6 MeV respectively and for carbon and oxygen the impact energies are up to $\sim$300 MeV. The additional physics expected at these energies, implies that a more detailed analysis of the particle erosion is required. In addition, because of the higher energies, especially from the other elements, the penetration distance would appear to exceed the minimum thickness required purely on the basis of material erosion considerations. It is therefore necessary to estimate the expected stopping distances and therefore the impact on any design decisions for particle bombardment shielding.

\begin{table}[ht!]
\begin{center}
     \begin{tabular}{| p{2cm} | p{2cm} | p{2cm}| p{2cm}| p{2cm}|} \hline
    \textbf{Speed} & {Hydrogen}  & {Helium} & {Carbon} & {Oxygen} \\ \hline
    0.05c & 1.17 & 4.66 & 13.99  & 18.63 \\ \hline
    0.1c   & 4.69 & 18.64 & 55.94 & 74.52  \\ \hline
0.15c   & 10.56 & 41.94  & 125.87 & 167.68 \\ \hline
0.2c   & 18.78 & 74.57 & 223.78  & 298.09 \\ \hline
0.25c   & 6.99 & 116.52 & 349.65 & 465.77  \\ \hline
            \end{tabular}
     \caption{\textbf{Collision energy (MeV) of impacting elements in interstellar medium as a function of speed of light.}}
\end{center}
 \end{table}

\hspace*{5in}

The loss of particle energy per unit distance is known as the stopping power $S(E)$ and is given by the kinetic energy $E_k$ of the particle and the distance $x$ that the particles move through.

\begin{equation}
S(E)=-\frac{dE}{dx}
\end{equation}

The stopping power can be expressed as a mass stopping power MeV/g/cm\textsuperscript{2} which would be $S(E)/\rho$, but in this work we will just use the stopping power and express it in units of mm given the small scale of the starshot probe, so that stopping power is MeV/mm. 

As a charged particle enters a material, it will gradually lose kinetic energy with distance until it reaches a maximum known as the Bragg Peak, and then the energy of the particle will drop to zero as it is fully deposited into the material. Typically, a $\sim$ 1 MeV charged particle will undergo around $\sim 10^5$ interactions prior to losing all of its kinetic energy, through both elastic and inelastic interactions. The mean range of the charged particle with initial kinetic energy $E_o$, into the material, is then calculated from the following integral as the Continuous Slowing Down Approximation (CSDA):

\begin{equation}
\Delta x = \int_o^{E_0} \frac{1}{S(E)} dE
\end{equation}

Where $S(E)$ is known as the stopping power of the target absorber material (in this case the spacecraft shielding material).

For the Starshot probe, we are interested in firstly calculating the stopping power for various charged particles and then computing the mean range of penetration into the material, and then to see if this range exceeds the defined thickness estimated from dust impacts.

The interaction of the charged particle with the impacting material will depend on the Coulomb interactions with the electrons in the material. For the absorbing medium, it is only the electron density that is important. It is assumed that the incoming charged particle has sufficiently high energy that it has lost its electrons. In addition, there are other effects to take into account such as quantum mechanics, relativistic velocities, excitation and ionisation, radiation losses such as Bremsstrahlung (>20 MeV) and even nuclear reactions (>100 MeV). This is all correctly accounted for in the stopping power. 

A good formulation of this is given by the Bethe equation which is derived from quantum mechanics and we adopt here only the leading order term $\sim$$z^2$ from perturbation theory. Higher order corrections to $\sim$$z^3$ and $\sim$$z^4$ have also been derived but we do not consider those here. It also takes into account relativistic effects and assumes that the incoming charged particle is much more massive than the electrons of the absorber, which is why we cannot use this equation for the impact of electrons onto the absorber, since they would have equal mass. The equation depends quadratically on the velocity and charge of the incoming particle $\sim z^2/ \beta^2$, but the energy loss is independent of the mass of the incoming particle. Originally derived by Niels Bohr in 1913 \cite{bohr}, and then re-derived by Hans Bethe in 1930 and 1932 \cite{bethe}, in particular to take account of relativistic quantum mechanics, it is given by \cite{lai, burrel}:

\begin{equation}
-\frac{dE}{dx}=\frac{4\pi n_e z^2 e^4 }{m_e c^2\beta^2 (4\pi \epsilon_o)^2}\Bigg[ln \Big(\frac{2m_e c^2 \beta^2}{I(1-\beta^2)} \Big)-\beta^2-log(I) \Bigg]
\end{equation}

Where $z$ is the atomic number of the heavy particle, $e$ is the electron charge, $n_e$ is the number of electrons per unit volume in the medium, $m_e$ is the electron rest mass, $c$ is the vacuum speed of light, $I$ is the mean excitation energy of the medium, $\beta$ is the speed of the particle relative to light. The negative energy change in Equation (9) reflects the deceleration of the charged particle with distance. As can be seen in Equation (9), the stopping power increases until the Bragg Peak, and then beyond that maximum it decreases approximately with inverse velocity squared. In the circumstance of low energy or low velocity impacts where $v<<c$ Equation (9) reduces to the non-relativistic version

\begin{equation}
-\frac{dE}{dx}=\frac{4\pi n_e z^2 e^4 }{m_e v^2 (4\pi \epsilon_o)^2}\Bigg[ln \Big( \frac{2m_e v^2}{I} \Big) \Bigg]
\end{equation}

The Ionisation potential (also called mean excitation energy) is the minimum amount of energy required to remove the most loosely bound electron from an atom. For a rough approximation ($\sim10-20\%$), it can be given by 

\begin{equation}
I\approx 9.1Z(1+1.9Z^{-2/3}) eV
\end{equation} 

And when substituted into Equation (9) or (10) yields the Bethe-Bloch equation. Note that Equation (11) is not accurate for compounds. Alternatively, empirical results can be used to estimate the ionization potential as a function of the atomic number of the absorber materials as follows \cite{turner}:

\begin{itemize}
\item For Z = 1 (e.g. hydrogen); $I \approx$ 19 eV
\item For 2 $\le Z \le 13$; $I \approx (11.2 + 11.7Z)$ eV
\item For Z > 13; $I \approx (52.8 +  8.71Z)$ eV
\end{itemize}

However, with access to exact values for the ionisation potential \cite{nist} we do not need to use this approximation. For this simple first order analysis we will restrict our assessment to Coulomb interactions only, and this approximation is known as Linear Energy Transferred (LET). The stopping power can then be calculated.

\begin{figure}[htb]
\includegraphics[width=6in]{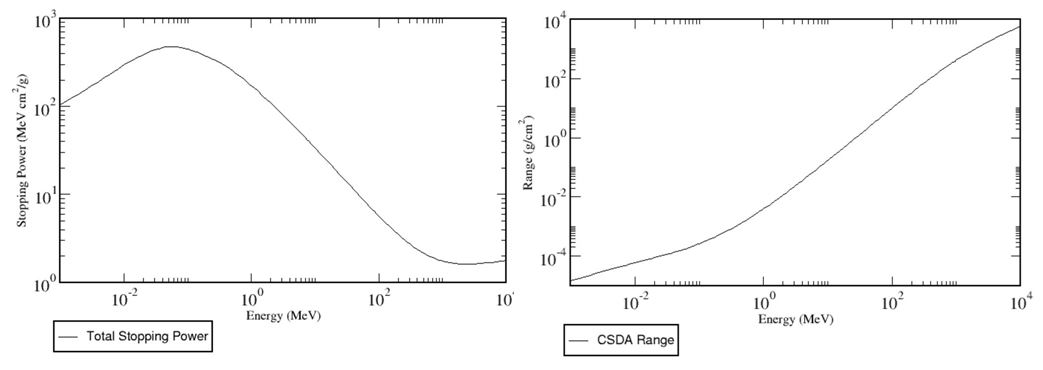}
\centering
\caption{\textbf{Calculations of Stopping Power and Range for Aluminium \cite{nist, nist2}.}}
\end{figure}

\begin{figure}[htb]
\includegraphics[width=6in]{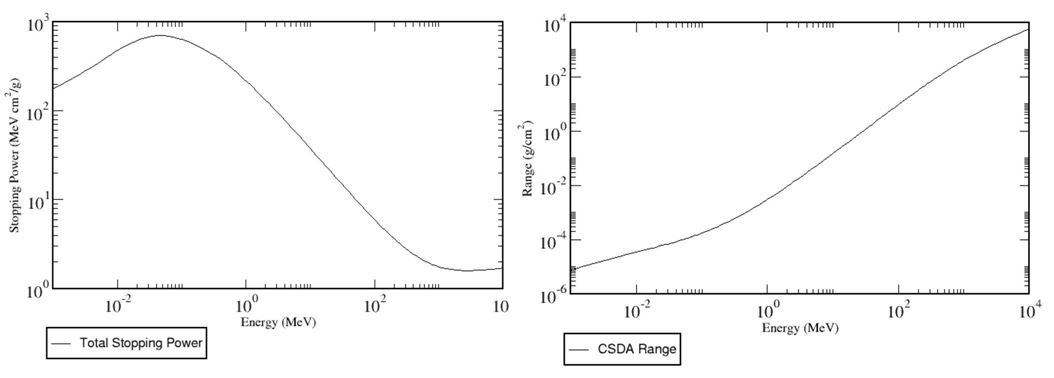}
\centering
\caption{\textbf{Calculations of Stopping Power and Range for Berryllium \cite{nist, nist2}.}}
\end{figure}

The electron number density is given as a function of the Avogadro number $N_A=6.022 \times 10^{23}$ mol\textsuperscript{-1}, the atomic number $Z$, the material density $\rho$, the relative atomic mass $A$ as follows:

\begin{equation}
n_e=\frac{N_A \rho}{A}
\end{equation}

Figures 2 and 3 show the stopping power and range estimated for Aluminium and Berryllium. Table 11 shows the values assumed for the different materials modelled in this work. The table also shows the results for the stopping power and range of penetration into the absorber material, for a $\sim$10 MeV ($\sim$0.15c) impacting proton. For Lithium, Berryllium, Boron and Graphite the derived ranges are much less than that derived from dust erosion as was shown in Tables 8-9.

\begin{table}[ht!]
\begin{center}
     \begin{tabular}{| p{3cm} | p{1.6cm} | p{1.4cm}| p{2cm}| p{2cm}| p{1.7cm}| p{1.5cm}|} \hline
    \textbf{Material} & {Rel.Atomic Wt (g/mol)}  & {Z/A} & {Electron Number Density (/m\textsuperscript{3})} & {Mean Excitation Energy ($ev$)} & {Stopping Power (MeV/mm)} & {Range (mm)}\\ \hline
    Lithium $Z=3$ & 6.9411 & 0.4322 & 4.598$\times10^{28}$  & 40 & 2.12  & 2.59\\ \hline
    Berryllium $Z=4$  & 9.0122 & 0.4438 & 1.236$\times10^{29}$ & 63.7 & 7.0 & 0.79 \\ \hline
Boron $Z=5$   & 10.8112 & 0.4625  & 1.392$\times10^{29}$ & 76 & 8.93 & 0.62  \\ \hline
Graphite $Z=6$   & 12.0105 & 0.4995 & 1.153$\times10^{29}$  & 78 & 9.21 & 0.60  \\ \hline
Aluminium $Z=13$  & 26.9816 & 0.4818 & 6.026$\times10^{28}$ & 166  & 9.25 & 0.63 \\ \hline
            \end{tabular}
     \caption{\textbf{Calculations of Electronic Stopping Power and results for a 10 MeV impacting proton.}}
\end{center}
 \end{table}

\subsection[Spacecraft Power]{Spacecraft Power}

Finally, it has been suggested that with the $\sim$0.3 J/sm\textsuperscript{2} energy flux during the flight this presents an opportunity for providing in-situ power for the spacecraft, for which some have speculated may be adapted for the application of deep space communications. It is worth exploring this possibility assuming that the incoming energy flux $\phi$ is a function of the generated power $P$ and frontal surface area $A_o$ as $\phi=P/A_o$.

If we assumed a frontal surface radii equivalent to the $\sim$1 mm 'Jerry' or $\sim$10 mm 'Tom' design geometries discussed in this paper, then this would lead to a potential power output (assuming 100\% energy conversion efficiency) of only $\sim1 \mu$W and $\sim100 \mu$W respectively. These estimates for $\mu$W power levels are consistent with estimates by Draine \cite{draine} who suggests the use for electrical power provision. Although this may be useful to power some on-board systems, using such a low power level for the purposes of long-range ($\sim$LY) antenna communications seems implausible. 

That said, as the spacecraft entered the target stellar target system, it is anticipated that this energy flux would increase and so may also the potential power levels and therefore the range of applications. This would require separate analysis to the cruise phase of the interstellar flight. We can however specify the energy flux levels that would be required to give rise to specific power levels, assuming the geometries of 'Jerry' and 'Tom' as above. In order for the 'Jerry' model to produce a power comparable to around 1 W, it would need to be impacted by an incoming energy flux of order $\sim$300,000 J/sm\textsuperscript{2}. Similarly, for the 'Tom' model to produce a 1 W comparable power level would require an incoming energy flux of  order $\sim$3,000 J/sm\textsuperscript{2}. Of course, such high energy fluxes would also lead to massively increased particle bombardment shielding requirements in order to account for the increased erosion rate, which may then go beyond the design space of Project Starshot.

In order for the currently calculated interstellar medium energy flux of $\sim$0.3 J/sm\textsuperscript{2} to give rise to a potential useful transmitter power output of 1 W, would require a front surface area on the spacecraft to be of order $\sim$0.3 m\textsuperscript{2} which corresponds to a frontal radius exceeding $\sim$300 mm. A proper trade study is required to estimate the optimum surface area for given levels of energy flux as a function of shielding mass requirements.

It goes without saying that if the Starshot probe had a larger sail area than the small systems considered in this study, then this would also increase the power potential. This wasn't considered in the above analysis because this author is assuming that after the acceleration phase that the sail is ejected, and only the probe continues onto the target destination. But if the sail was included this would change the potential power output. The current baseline sail area for the Starshot mission includes using a sail with a diameter of 4 m. For an assumed energy flux of $\sim$0.3 J/sm\textsuperscript{2} this would suggest a potential power production of around $\sim3.7$ W, and over the 21.5 lifetime flight of the mission would be equivalent to an incoming energy requirement on the sail surface of $\sim$2,471 MJ. Alternatively, if the sail was only opened once it reached the target destination, and for a duration of 1 year as it encountered the target, this would be equivalent to an energy requirement of $\sim$115 MJ.

Although, as the starshot probe reaches the target destination it is expected that the energy flux will increase significantly from the assumed $\sim$0.3 J/sm\textsuperscript{2}. For example, if the energy flux was much higher at values of $\sim$3 J/sm\textsuperscript{2} ($\times10$), $\sim$30 J/sm\textsuperscript{2} ($\times100$) or $\sim$300 J/sm\textsuperscript{2} ($\times1000$), then for a 4 m diameter sail this would give rise to power levels of $\sim$ 37 W, $\sim$377 W and $\sim$3,770 W respectively. Clearly, it is important to the mission of Project Starshot to accurately estimate the expected energy flux both during the cruise phase of the mission (for the average power) and during the encounter with the target star (for the peak power). Discussions within the Breakthrough Starshot team indicate that for successful data retrieval, there will be a likely requirement fora 10 milli-W average power and a 10 kW peak power. The latter is suggestive of a requirement for an energy flux at the target encounter approximately 1,000$\times$ that currently estimated for the interstellar medium during the cruise phase.

\section[Conclusions]{Conclusions}

In this paper we have attempted to build a preliminary model for the prediction of particle bombardment on the frontal end surface of the gram-scale interstellar starshot probe as it moves through the interstellar medium. We have modelled particle distributions of expected typical, mean and largest dust masses. Due to the high velocity of the probe, 60,000 km/s, the impacts due to dust erosion will need to be taken into account and for the models examined in this paper will result in a frontal area temperature on the spacecraft of $\sim$135.2 K assuming $A_o/A \sim$0.1.

If a bombardment shield was to be utilised, constrained within a total spacecraft mass of 1 gram, we estimate that it would have a mass of order $\sim$0.01 - 0.05 g, depending on the material chosen. This would be between $\sim1 - 5\%$ of the total mass, assuming a spacecraft mass not exceeding 1g. The inclusion of this shield would obviously affect the payload mass fraction possible for the mission. On the basis of this analysis, a material shielding of order $\sim$1.4 - 3 mm thickness would likely be sufficient to mitigate against any dust impacts. This is a for a cylindrical spacecraft geometry with radii $\sim$1 mm and length $\sim$5 mm on a 21.5 year mission duration.

However, this estimate only considers the physics effects in isolation and we must also take into account a design margin. So instead it may be more sensible to consider the accumulated effects to arrive at an estimate for the overall design recommendation. From the impact model estimates of section 2.1 we assumed a penetration depth of up to 1 mm. For the erosion rate estimates of section 2.3 we calculated a minimum shielding thickness in the range 1.47 mm and a maximum up to 3.01 mm. For the charged particle analysis of section 2.4 we calculated a minimum particle range of 0.6 mm and a maximum of up to 2.59 mm. It can be concluded therefore that a recommended shielding thickness may be at a minimum of 3.07 mm but may be as large as 6.6 mm, depending on material choice. This would need to be examined further.

Using a spacecraft shielding of too high a density will lead to excessively massive shields, or using too low a shielding density will lead to too long a spacecraft shield, which takes the design away from plausible concepts given the assumed mass constraint requirement, and also the desire to maintain beaming power levels within a few $\sim$100s GW range. 

The results of this work highlight the important design coupling between the choice of shielding material and the geometry (frontal and radiating surface area) of the spacecraft which have to be taken into consideration as a part of systems engineering when developing vehicle concepts for Project Starshot. In this work we have not considered geometries alternative to the cylinder (i.e. torus, sphere) but these could be considered by others in future analysis. 

Some analysis has also been conducted on the effects of charged particles such as electrons and ions, which need to be calculated separately as a charged particle transport problem, as a solution to the Bethe equation for stopping power. It is expected that the inclusion of charge particle effects will lead to an increase in the shielding mass and thickness requirement and this is particularly the case when considering impacts by ionised hydrogen from the interstellar medium. A more detailed analysis will need to be conducted to properly characterise the shielding requirements to guarantee survivability of the probe. Particularly since there is the potential for incoming protons to produce bremsstrahlung photons deep below the surface of the material as studied by others \cite{lubin3}.

The analysis conducted in this paper on particle bombardments on an interstellar probe, does not take into consideration material ejected from larger objects such as asteroids or comets, although the statistical chance of an encounter is very low unless the mission is directed towards such an object.

A brief examination on the possibility for converting incoming energy flux at cruise speed of $\sim$0.3 J/sm\textsuperscript{2} into useful power for the purposes of long-range communications does not currently show potential for this idea with the two configuration geometries examined in this work. Although if one accepts a much larger surface area (e.g. 4 m diameter sail), the potential becomes more realisable. In addition, it is expected that this potential may be significantly improved when considering the energy flux on approaching the target stellar system, although an estimated energy flux of $\times$1,000 of the interstellar medium would be required to result in useful power outputs in the $\sim$kW range required for peak power transmission. An increased spacecraft mass and associated transmitter area would also increase this potential, but at the cost of increased beamer power for the mission. 

Finally, it should be self-evident that getting more data on the properties of the interstellar medium will be extremely valuable to our confidence in calculating the particle bombardment problem. This could be through enhanced data collection methods such as launching more missions like the NASA Stardust mission which deployed both a Dust Flux Monitor Instrument (DFMI) as well as the Stardust Sample Collection (SSC) system.  In addition, in principle it is possible to validate our models of dust and particle impacts using experimental facilities such as the energetic light gas guns used at the Experimental Impact Laboratory, NASA Johnson Space Centre. Although this facility is only limited to launch speeds of 8 km/s firing projectiles from as small as 1 $\mu$ m up to 5.56 mm in diameter. It would be constructive to initiate a discussion with such facilities about what it would take to get a relativistic gas gun operational, in terms of cost and logistics. Even if such a facility was nowhere near the operational mission regime of Project Starshot (0.2c), extrapolating from that experimental data would still be of value towards validating the mission as credible; as opposed to having no experimental validation at all. This will also be useful in calibrating any numerical simulations of the problem.

\section[Acknowledgements]{Acknowledgements}
The author thanks David H Johnson for assistance with the charged particle calculations using a script written in Python and based on the algorithms of Swift and NcNaney \cite{swift1}. The author would like to thank Phil Lubin and Pete Klupar for discussions which informed this paper early on. Any errors or omissions however are the sole responsibility of the author. Elements of this work was originally presented at the NASA Lyndon B Johnson Space Center Thermal Design Branch, Houston, USA, February 2017; the 1st Foundations of Interstellar Studies workshop, City Tech, New York, June 2017; and also at the 60th Cospar meeting, Pasadena, USA, July 2018. A thank you also to Eric Malroy and the NASA Stardust laboratory for facilitating a tour of the facilities and interstellar dust results of the Stardust mission. We acknowledge the Breakthrough Initiatives and Pete Worden in supporting this research. This work was completed whilst the author served on the Project Starshot Advisory Committee but the opinions expressed are his own.

\end{document}